\input phyzzx
\newcount\mongocount
\mongocount=1
\def\Figure#1#2#3{
      \vbox to #3in{\hsize=#2in
        \vfil
         \includegraphics{#1}
    }
}
\def\figcap#1#2{
\vtop{\tenpoint\singlespace
\hsize=#1in\smallskip\noindent Figure\ \ \the\mongocount.\ \  #2
\global\advance\mongocount by 1\bigskip}}
\def\mongofigure#1#2#3#4#5{\centerline{\Figure{#1}{#2}{#3}
\figcap{#4}{#5}}}

\hoffset=0.375in
\def\lmc{{\rm LMC}}

\def\kms{\rm km\,s^{-1}}

\def\sn{{\rm SN}}
\def\max{{\rm max}}

\font\bigfont=cmr17
\centerline{\bigfont Upper Limit to the}
\smallskip
\centerline{\bigfont Distance to the Large Magellanic Cloud}
\bigskip
\centerline{{\bf Andrew Gould}\foot{Alfred P.\ Sloan Foundation Fellow} and
\bf Osamu Uza}
\bigskip
\centerline{Dept of Astronomy, Ohio State University, Columbus, OH 43210}
\smallskip
\centerline{E-mail:  gould,uza@astronomy.ohio-state.edu}
\bigskip
\singlespace
\centerline{\bf ABSTRACT}

	We remeasure the ``light echo'' times to the near and far side
of the ring around SN 1987A using the method of Gould (1995) and the
new reductions of the original UV spectra by Sonneborn et al.\ (1996).
Under the assumption that the ring is circular, we obtain an upper limit to 
the distance to the Large Magellanic Cloud $\mu_\lmc < 18.37\pm 0.04$.  
This result is very close to Gould's original value, and contradicts claims 
that the new spectra measurements lead to a significant increase in the 
distance estimate and a substantial increase in the statistical errors.  
We present new evidence (partly provided by A.\ P.\ S.\ Crotts) that the ring 
is intrinsically elliptical, with axis ratio $b/a\sim 0.95$.  For an elliptical
ring, $\mu_\lmc < 18.44\pm 0.05$.  We elaborate on Gould's original argument 
that the upper limit is robust, i.e., that it is only weakly sensitive to
unmodeled aspects of the ring such as its hydrodynamic properties and
ionization history.

Subject Headings:  galaxies: distances and redshifts -- Magellanic Clouds --
supernovae: individual (SN 1987A)
\bigskip

\chapter{Introduction}

	The distance to the Large Magellanic Cloud (LMC) has been 
controversial.  A low value ($\mu_\lmc = 18.28\pm 0.13$) is obtained from RR 
Lyraes calibrated by statistical parallax (Layden et al.\ 1996; 
Popowski \& Gould 1997), while
much higher values have recently been obtained from Hipparcos-based 
calibrations of RR Lyraes by Reid (1997) ($\mu_\lmc=18.65\pm 0.1$) and
by Gratton et al.\ (1997) ($\mu_\lmc=18.63\pm 0.06$) and of Cepheids by
Feast \& Catchpole (1997) ($\mu_\lmc=18.70\pm 0.10$).  

	The light curves of fluorescent UV line emission from the ring
around SN 1987A permit an independent determination of the distance to the
LMC (Panagia et al.\ 1991; Gould 1994, 1995; Sonneborn et al.\ 1996).  Stripped
to its essence, the method can be understood as follows.  From simple
geometric considerations the light curve should have two cusps, the first
corresponding to the excess time for light to travel from the supernova to 
the near side of the ring to Earth (relative to direct travel from
the supernova to Earth), and the second corresponding to the excess light 
travel
time to the far side of the ring.  The cusps arise because the parabola of 
constant delay time is tangent to the ring at these two times.  The cusps
are clearly observable in the N~III] and N~IV] lines.  Assuming that the
ring is coplanar with the supernova, the light travel time across the apparent
minor diameter of the ring is then $(t_+ + t_-)$ where $t_\pm$ are the
times of the two cusps.  The apparent major and minor diameters of the
ring ($\theta_\pm$) have been measured by Plait et al.\ (1995) who find
$$\theta_+ = 1.\hskip-2pt ''716 \pm 0.\hskip-2pt ''022,\qquad 
\theta_- = 1.\hskip-2pt ''242 \pm 0.\hskip-2pt ''022 \
.\eqn\thetapm$$ 
If the ring is assumed to be circular, one can therefore immediately derive
a distance
$$D_{\rm SN} = {c(t_+ + t_-)\over \theta_+}.\eqn\naivedist$$
Since the error in $\theta_+$ is small ($\sim 1\%$),  equation
\naivedist\ has the potential to yield a remarkably precise distance provided
that $t_\pm$ can be measured equally well.

If the ring is assumed to be circular, then there are two independent
methods of estimating its inclination, $i$, one from the ratio of the
axes, $\eta_\theta \equiv \theta_-/\theta_+$, and the other from the ratio
of the delay times, $\eta_t \equiv t_-/t_+$,
$$i_\theta = \cos^{-1}\eta_\theta, \qquad i_t = {\pi\over 2} - 
2\tan^{-1}\eta_t^{1/2}.\eqn\inclines$$
In practice therefore, rather than directly applying equation \naivedist,
one should appropriately weight both of these estimates of $i$
to determine the angular size and inclination of the ring.
A more serious potential problem with equation \naivedist\ is that it ignores 
various physical effects that could affect the measured values of $t_\pm$ other
than simple light travel times.  We address these in \S\ 4 and \S\ 5.
Nevertheless, equation \naivedist\ gives a good representation of the 
underlying simplicity and geometric nature of the method.

\chapter{Previous Work}

	To use the supernova ring to measure the distance to the LMC, one
must both measure the distance to the ring and also
determine the distance of the
center of mass of the LMC relative to the ring.  Since these are logically
distinct, we review them separately.

\section{LMC Distance Relative to the Ring}

	Jacoby et al.\ (1992) were the first to point out that at the position
of SN 1987A, $\sim 1\,$kpc from the center of the LMC bar, the plane of
the LMC is $\sim 500\,$pc in front of the LMC center of mass, assuming a
standard LMC inclination angle of $27^\circ$ (Bessel, Freeman, \& 
Wood 1986).  In principle, the supernova could be anywhere along the line
of sight, but Gould (1995) argued that since SN 1987A is a Pop I object, its 
most likely position is in the plane of the LMC.  While fairly compelling, this
argument is nevertheless wrong.  Xu, Crotts, \& Kunkel (1995) 
mapped the three dimensional
structure of the dust in front of SN 1987A using light echos.  If the supernova
were in the plane, one would expect to find large quantities of dust
within $\sim 100\,$ pc of the supernova and progressively lower densities
farther away.  In fact, Xu et al.\ (1995) find almost no material within 
100 pc, while
the largest concentration of dust, the huge LH90 cloud, lies 490 pc in front
of SN 1987A, implying that the plane of the LMC is $\sim 500$ pc in front of 
the center of mass. Since the supernova is $\sim 500$ pc behind the plane, we 
conclude that the supernova and the LMC are at the same distance,
$\mu_\lmc=\mu_\sn$.

\section{Distance to the Ring}

	Panagia et al.\ (1991) measured the light curves from four ions
(N~III], N~IV], N~V, and C III]).  By comparing these with models, they
found $t_+ = 413\pm 24\,$days and $t_- = 83\pm 6\,$days.  They combined
these with early measurements of $\theta_\pm$ by Jakobsen et al.\ (1991)
(which are $\sim 3\%$ smaller than eq.\ \thetapm) to obtain a distance
of $\mu_\sn=18.55\pm 0.13$.  Dwek \& Felten (1992) argued that the model
adopted by Panagia et al.\ (1991) was inappropriate for a ring geometry.
Gould (1995) developed the suggestion of Dwek \& Felten (1992) into a
systematic mathematical treatment. He then reanalyzed the data as presented in
figure form by Panagia et al.\ (1991) and derived new delay times
$t_+=390.0\pm 1.8\,$days, $t_-=75.0\pm 2.6\,$days.  Using these and the new
measurements of the ring size by Plait et al.\ (1995) (see eq.\ \thetapm),
he obtained a distance estimate $\mu_\sn=18.350\pm 0.035$.  Gould (1995)
restricted his analysis to the nitrogen ions, arguing that the carbon ion
could not be put on a common basis with the other three.  He also argued
that since N~V is a permitted line, it could well be optically thick
which would vitiate the basic analysis.  Nevertheless, Gould (1995) included
N~V in the analysis for completeness since, as he showed, it changed neither
the best fit nor the error bars significantly.

	Sonneborn et al.\ (1996) have now re-reduced the data originally used
by Panagia et al.\ (1991).  They obtain a new estimate for $t_+$ and $t_-$
(or rather $t_{\rm max}$ and  $t_{\rm rise}$
which are not exactly the
same things, see \S\ 6) based
on a high-order polynomial fit to the N~III] light curve,
$$t_{\rm max} = 399\pm 15\,{\rm days},\quad  t_{\rm rise} 
= 83\pm 4\,{\rm days}\qquad
({\rm Sonneborn}\ {\rm N}\ {\rm III])}.\eqn\sonntptm$$
Based on these values including their ``more realistic errors'', they scale
the results of Gould (1995) to obtain
$$\mu_\sn = 18.43\pm 0.10\qquad
({\rm Sonneborn}\ {\rm N}\ {\rm III])}.\eqn\sonndis$$

	Although Sonneborn et al.\ (1996) do not regard this as a definitive 
result (they defer a more detailed treatment to future work), it has encouraged
some to believe that the conflict between the supernova-ring distance and
other estimates of the distance to the LMC is substantially less severe
than originally claimed by Gould (1995).  See, for example,
Feast \& Catchpole (1997).  We therefore investigate the
effect of applying Gould's (1995) method to the
new reductions by Sonneborn et al.\ (1996).

\chapter{Redetermination of the Ring Distance}

\FIG\one{
Previous data and light-curve fit for N~III] emission from SN 1987A
(based on
Fig.\ 3 of Gould 1995) with data taken from Panagia et al.\ (1991).
Intensity is in units of
$10^{-14} \rm erg\ cm^{-2}\ s^{-1}$.  Shown are data points from GSFC
({\it circles}) and VILSPA ({\it crosses}).
The curve is the best fit to the data using eq.\ (2.3) of Gould (1995) with
$t_-=83\,$days, $t_+=390\,$days and assuming a rise time of $t_*=1\,$day.
The remaining parameters are the amplitude $A=67$, and the decay time  
$\tau=276\,$days.  VILSPA data have been multiplied by an amplitude correction 
factor $Q=0.88$ which minimizes scatter of the fit.  
}
\midinsert
\mongofigure{ps.niiix}{5.5}{5.4}{5.5}
{
Previous data and light-curve fit for N~III] emission from SN 1987A
(based on
Fig.\ 3 of Gould 1995) with data taken from Panagia et al.\ (1991).
Intensity is in units of
$10^{-14} \rm erg\ cm^{-2}\ s^{-1}$.  Shown are data points from GSFC
({\it circles}) and VILSPA ({\it crosses}).
The curve is the best fit to the data using eq.\ (2.3) of Gould (1995) with
$t_-=83\,$days, $t_+=390\,$days and assuming a rise time of $t_*=1\,$day.
The remaining parameters are the amplitude $A=67$, and the decay time  
$\tau=276\,$days.  VILSPA data have been multiplied by an amplitude correction 
factor $Q=0.88$ which minimizes scatter of the fit.  
}
\endinsert
\FIG\two{
New data and light-curve fit for N~III] emission from SN 1987A
with rereduced data taken from Sonneborn et al.\ (1996).
Intensity is in units of $10^{-14} \rm erg\ cm^{-2}\ s^{-1}$.  
Shown are data points from short ($\sim 80\,$minute) 
({\it solid squares}) 
and long ($\sim 250\,$minute) ({\it open circles}) 
exposures.
The solid curve is the best fit to the data using 
eq.\ (2.3) of Gould (1995) with
$t_-=88\,$days, $t_+=381\,$days and assuming a rise time of $t_*=1\,$day.
The remaining parameters are the amplitude $A=86$, and the decay time  
$\tau=279\,$days.  The dashed curve is an $n=9$ order polynomial fit to
the data (following Sonneborn et al.\ 1996).
}
\midinsert
\mongofigure{ps.niii_newx}{5.5}{5.4}{5.5}
{
New data and light-curve fit for N~III] emission from SN 1987A
with rereduced data taken from Sonneborn et al.\ (1996).
Intensity is in units of $10^{-14} \rm erg\ cm^{-2}\ s^{-1}$.  
Shown are data points from short ($\sim 80\,$minute) 
({\it solid squares}) 
and long ($\sim 250\,$minute) ({\it open circles}) 
exposures.
The solid curve is the best fit to the data using 
eq.\ (2.3) of Gould (1995) with
$t_-=88\,$days, $t_+=381\,$days and assuming a rise time of $t_*=1\,$day.
The remaining parameters are the amplitude $A=86$, and the decay time  
$\tau=279\,$days.  The dashed curve is an $n=9$ order polynomial fit to
the data (following Sonneborn et al.\ 1996).
}
\endinsert
\FIG\three{
New data and light-curve fit for N~IV] emission from SN 1987A
with rereduced data taken from Sonneborn et al.\ (1996).
Similar to Fig.\ \two, except
$t_-=66\,$days, $t_+=378\,$days, $A=53$, and $\tau=180\,$days.  
Only one type of data point is shown because all exposures are of similar
duration.  
}
\midinsert
\mongofigure{ps.niv_newx}{5.5}{5.4}{5.5}
{
New data and light-curve fit for N~IV] emission from SN 1987A
with rereduced data taken from Sonneborn et al.\ (1996).
Similar to Fig.\ \two, except
$t_-=66\,$days, $t_+=378\,$days, $A=53$, and $\tau=180\,$days.  
Only one type of data point is shown because all exposures are of similar
duration.  
}
\endinsert
\FIG\four{
Goodness of fit ($\chi^2$ relative to its minimum value)
for the light curves of N~III] ({\it solid}) and
N~IV] ({\it dashes}), as a function of the
time of the first caustic, $t_-$.  The sum of the two curves is shown
as a bold line.  $t_+$ is held fixed at 381 and 378 days for N~III] and
N~IV], respectively.
To be compared with Fig.\ 1 of Gould (1995).
}
\midinsert
\mongofigure{ps.1cuspx}{5.5}{5.4}{5.5}
{
Goodness of fit ($\chi^2$ relative to its minimum value)
for the light curves of N~III] ({\it solid}) and
N~IV] ({\it dashes}), as a function of the
time of the first caustic, $t_-$.  The sum of the two curves is shown
as a bold line.  $t_+$ is held fixed at 381 and 378 days for N~III] and
N~IV], respectively.
To be compared with Fig.\ 1 of Gould (1995).
}
\endinsert
\FIG\five{
Goodness of fit ($\chi^2$ relative to its minimum value)
for the light curves of N~III] ({\it solid}) and
N~IV] ({\it dashes}), as a function of the
time of the second caustic, $t_+$.  The sum of the two curves is shown
as a bold line.  $t_-$ is held fixed at 88 days for N~III] and 66 days for
N~IV.  To be compared with Fig.\ 2 of Gould (1995).
}
\midinsert
\mongofigure{ps.2cuspx}{5.5}{5.4}{5.5}
{
Goodness of fit ($\chi^2$ relative to its minimum value)
for the light curves of N~III] ({\it solid}) and
N~IV] ({\it dashes}), as a function of the
time of the second caustic, $t_+$.  The sum of the two curves is shown
as a bold line.  $t_-$ is held fixed at 88 days for N~III] and 66 days for
N~IV.  To be compared with Fig.\ 2 of Gould (1995).
}
\endinsert

	Except where otherwise noted, we follow the analysis given (and
justified in some detail) by Gould (1995).
We model the light curves according to the prescription given by his
equation (2.3).  This assumes that each point on the ring
responds promptly (1 day rise time) to the EUV blast, and then exponentially
decays.  (We relax the assumption of prompt response below.)
We restrict attention to N~III] and N~IV] (see \S\ 2.2) and to data
from the first 700 days (see Gould 1995).  We
slightly modify Gould's (1995) procedure for establishing error bars.  Gould
assumed equal error bars for all points and normalized these to make
$\chi^2/dof\equiv 1$.  However, Sonneborn et al.\ (1996) give exposure times
for each data point. For N~III] (but not N~IV]), these fall into two classes:
48 with exposure times of $\sim 300$ minutes and 15 with  exposure times of 
$\sim 80$ minutes.  
We find that the latter have about twice the scatter about the
best fit curve as the former.  We therefore set the errors on the
shorter exposures at twice the value for the longer exposures before
normalizing them by $\chi^2_{\rm min}= 59$  (corresponding to 63 points less
4 fitting parameters, i.e., $t_\pm$, amplitude, and decay time).  
	
	Figures \one--\three\ show best fits for the light curves together
with the N~III] and N~IV] data upon which they are based. Figures \two\ and
\three\ show the best fit curves from the current analysis of the Sonneborn 
et al.\ (1996) data for N~III] and N~IV]. 
(Fig.\ \two\ also shows a 9th order polynomial fit which is discussed in
\S\ 6).  For
comparison we show in Figure \one\ the fit from Gould (1995) for N~III]
together with the data (from Panagia et al.\ 1991) on which it was based.

	Comparison of Figures \one\ and \two\ shows that there are many more
points in the former (143 vs.\ 63).  Part of the reason is that the circles and
crosses represent separate reductions of the same data by the GSFC and VILSPA
stations respectively.  Gould (1995) did not realize that the 45 VILSPA points
were redundant and so incorrectly included them in his analysis.  However,
that still leaves the problem of why Panagia et al.\ (1991) show 98 points
(circles) while Sonneborn et al.\ (1996) show only 63.  Since Panagia et al.\
(1991) have not published their data in tabular form, we cannot resolve this
issue.  We assume that the new reductions by Sonneborn et al.\ (1996) are
correct.  Figure \two\ also shows a modest shift in the second cusp from
390 to 381 days for N~III].  N~IV] shows an almost identical shift.  See below.

	Figure \four\ shows $\Delta\chi^2(t_-)\equiv
\chi^2(t_-)-\chi^2_{\rm min}$, where $\chi^2(t_-)$ is the value of $\chi^2$
for the best fit subject to constraining $t_-$ to that value and $t_+$ to
381 and 378 days for N~III] and N~IV], respectively.  
The minimum value (defined above to be equal to the number of data 
points minus the number of parameters) is subtracted out to allow easy
comparison of different curves.  Figure \four\ shows N~III] and N~IV] 
separately, as well as their sum.  Figure \five\ shows $\Delta\chi^2(t_+)$
which is defined similarly, with $t_-$ held fixed at 66 days and 88 days, i.e.,
at its best-fit values as shown in Figure \four.

	When we use these measurements of $t_\pm$ to evaluate $\mu_\sn$,
we will directly employ the $\chi^2$ values shown in Figures \four\ and \five.
However, for purposes of discussion, it is useful to state the
results in terms of best fits.  To this end, we follow Gould (1995) and 
estimate the best fit as the center of the ``$2\,\sigma$ interval'' 
($\Delta\chi^2<4$) and the error as 1/4 of the width of this interval.  We
then find $t_-=87.8\pm 2.7\,$days and $t_+ =380.7 \pm 6.3\,$days for N~III], 
and $t_-=65.6\pm 5.6\,$days and $t_+= 377.8\pm 8.6\,$days for N~IV].
For the combined fit, we find
$$t_- = 80.5\pm 1.7\,{\rm days},\qquad t_+ = 378.3 \pm 4.8\,{\rm days}.
\eqn\tpmnew$$

	Before using these results to measure the distance to the ring, we
briefly comment on the nature of the changes relative to Gould's (1995)
determination.  The most striking difference is the increase in the error
in $t_+$ based on N~III] from 3.2 to 5.6 days, a factor of 1.75.  
A factor $\sim 1.5$ of this
is due to the reduced number of points (see above).  Most of the rest is due
to the fact that the lower quality points (squares) are concentrated near the
peak which adversely affects the accuracy of its determination.  The best 
fit values for $t_+$
are consistent for the two ions, but those for $t_-$ are not.  This is
basically the same situation found by Gould (1995) although the inconsistency
for $t_-$ is now less severe.  The overall best-fit values rose by
$\sim 5\,$days for $t_-$ and fell by $\sim 12\,$days for $t_+$.  
This opposing motion will imply that the estimate of the distance changes
very little, but that the
consistency between $i_t$ and $i_\theta$ is decreased.

	Our primary method of estimating the distance to the ring is to
assume that the ring is circular with unknown distance $D$, radius, $r$, and
angle of inclination, $i$.  For each distance, we then sum over all 
combinations of $r$ and $i$ and weight by the probability of obtaining
the observed values of $\theta_\pm$ and $t_\pm$ given their model values.  The 
model values are $\theta_+(D,r,i) = 2r/D$, $\theta_-(D,r,i)=2r\cos i/D$, and
$t_\pm(D,r,i) = (r/c)(1\pm\sin i)$.  The relative probability of a given
distance is then
$$\eqalign{P(D)& = \int d r\int d i \exp\biggl\{- {1\over 2}
\biggl[\chi^2[t_+(D,r,i)] + \chi^2[(t_-(D,r,i)] \cr  & +
\biggl[{\theta_+(D,r,i)-\theta_{+,\rm obs}\over \sigma_+}\biggr]^2 + 
\biggl[{\theta_-(D,r,i)-\theta_{-,\rm obs}\over \sigma_-}\biggr]^2 
\biggr]\biggr\},}\eqn\prob$$
where $\theta_{\pm,\rm obs}$ and $\sigma_\pm$ are given by equation
\thetapm, and $\chi^2[t_\pm(D,r,i)]$ are given by Figures \four\ and \five.
We then find,
$$\mu_\sn = 18.372\pm 0.035\qquad ({\rm all}\ {\rm data}).\eqn\alldata$$

\chapter{Alternative Interpretations}

	There are, however, several alternative viewpoints on how to treat
the data.  First, the two determinations of $t_+$ are quite consistent, but
the two determinations of $t_-$ are discrepant at the $4\,\sigma$ level.  
Gould (1995) argued that this probably arose from the fact that the ring
almost certainly deviates from the simple model that forms the basis of the
light curve analysis.  He showed that the determinations of $t_+$ were likely
to be robust in the face of these deviations but those of $t_-$ were not.
Thus, one might assume that the ring is circular, but ignore all information
about $t_-$.  The resulting distance is then
$D_\sn = c t_+/[\theta_+(1 + \sin i_\theta)]$, or
$$\mu_\sn = 18.29\pm 0.05\qquad ({\rm excluding}\ {t_-}).\eqn\excmin$$

	Second, it is not immediately obvious that the optical emission lines
used by Plait et al.\ (1995) to measure the size of the ring arise from the
same gas that generated the UV fluorescent emission used to measure $t_\pm$.
Plait et al.\ (1995) find an upper limit for the full width half maximum of 
the ring of $0.\hskip-2pt ''121$, i.e., 7\% of its major diameter.  
If the UV emission actually arose from
the inner edge of the ring, but the optical emission used to measure the
size of the ring came from the ring as a whole, then the UV light curve
would yield an underestimate of the light-travel time across the optical
ring diameter by up to 7\%, and so [from eq.\
\naivedist] {\it underestimate} the distance to the ring by the same amount.
In this case,
$$\mu_\sn < 18.53\pm 0.04,\qquad
({\rm optically}\ {\rm thick}\ {\rm N~III}/{\rm thin}\ {\rm O~III})
,\eqn\optthick$$
where the inequality reflects the fact that Plait et al.\ (1995) find only
an upper limit for the ring thickness.

We consider this scenario to be highly implausible, fundamentally because
O~III and N~III have similar ionization potentials so it is difficult to
see how the mean radii of their emission
distributions could be substantially different.
We note the following specific observational evidence against separate
emission.
First Plait et al.\ (1995) find that the best fit model for the O~III emission
is a crescent ring such as would be produced if the ring were optically thick
to the EUV blast and thus only the inner face were illuminated.  If this model
is correct, then only a thin film of gas would contain either N~III or O~III,
so the emission from the two ions would be cospatial.  Plait et al.\
(1995) cannot actually rule out a toroidal distribution for the [O~III]
emission, such as would be produced if the ring were optically thin to the
EUV blast.  

	However, even if the [O~III] emission is toroidal, it should still be 
cospatial with N~III].  The [O III] emission
was first observed 1278 days after the supernova core collapse, and thus
2.4 yr after the peak of fluorescent emission at $t_+=380\,$days.  
If the ring were in fact
optically thick to the EUV blast, so that only the inner face fluoresced in
N~III] and N~IV], then how did the O III in the rest of the ring
become ionized?  There are only two possibilities:
either it was ionized by a shock wave propagating through the ring from the 
inner face, or it was ionized by other wavelengths of UV radiation to which
the ring was optically thin.  There are two arguments against the shock-wave
hypothesis.  First, if such a turbulent process had proceeded across the 
ring in only 2.4 years,
then one would expect that in the next 2.4 years, the ring would show some
evidence of gas motions on the scale of its thickness.  To the contrary,
however, Plait et al.\ (1995) report that the late time behavior of the
ring over the following 2.7 years is simple fading.  Second, from the Plait 
et al.\ (1995) measurement of the ring thickness and the 2.4 year maximum
time for the shock propagation, we can infer a minimum shock speed of 
$10^4\,\kms$.
One would then expect a substantial imprint on the bulk expansion of the ring.
However, Crotts \& Heathcote (1991) measure an expansion velocity smaller
than this by a factor $10^{-3}$.  If the O III was ionized by other
wavelengths of UV radiation
to which the ring was optically thin, then this radiation should have also
generated N~III (which has a similar ionization potential) throughout the
ring.  Since there is vastly more nitrogen in the ring as a whole than there
is on the inner face, the centroid of the N~III] emission should then have been
close to the center of the optical ring (as was assumed in deriving 
eq.\ \alldata).  

	In brief, the fact that N~III (which Fig.\ \two\ shows  to have a 
very well defined light curve) and O~III have similar ionization potentials
implies that their physical distributions should be similar.  Equation 
\optthick\ therefore represents an extreme upper limit for a physically 
implausible scenario.

	Finally, the assumption that the ring is circular may not be valid.  
Gould (1994) showed that if the ring is elliptical, but if 
$i_t\sim i_\theta$ to within statistical errors, then the inferred distance
is overestimated by a factor ($1+0.4e^4$) where $e$ is the eccentricity.
Using equations \thetapm, \inclines, and \alldata, we find
$$i_\theta = 43.\hskip-2pt ^\circ 6\pm 1.\hskip-2pt ^\circ 3,\qquad
i_t = 40.\hskip-2pt ^\circ 5\pm 0.\hskip-2pt ^\circ 5.\eqn\ievals$$
These values are discrepant at the $2\,\sigma$ level, implying that 
Gould's (1994) theorem
does not strictly apply.  For the particular set of measurements
\thetapm\ and \alldata, the maximum distance occurs if the 
apparent minor axis of the ellipse is aligned with
the true minor axis, in which case equation \naivedist\ is modified to become 
$D_{\rm SN} = {c(t_+ + t_-)\cos i_\theta /\theta_-}$.  This yields
$$\mu_\sn = 18.44\pm 0.05,\qquad
({\rm aligned}\ {\rm ellipse}).\eqn\excthp$$
The axis ratio in this case would be 
$${b\over a}={\cos i_\theta\over\cos i_t} =0.95\pm 0.02\qquad 
({\rm ring}),\eqn\bovera$$
and the eccentricity would be $e\sim 0.3$.
Crotts, Kunkel, \& Heathcote (1995) showed that the 
three-dimensional structure of the double-lobed 
nebula (of which the ring forms a ``waist'') is close to axisymmetric, but
did not attempt to quantify the possible deviations from axisymmetry.
However, A.\ P.\ S.\ Crotts (1997, private communication) has now conducted
a new analysis of the Crotts et al.\ (1995) data and has generously made these
available in advance of publication.  He finds that the nebula is intrinsically
flattened in the sense that the shorter diameter is approximately aligned
with the apparent minor axis of the ring.  If the axis of the three-dimensional
structure is assumed to be inclined at $i=40^\circ\hskip-2pt .5$, then Crotts
finds that the axis ratio of the nebula is almost independent of distance
from the ring and has a mean value of
$${b\over a} =0.95\pm 0.02\qquad ({\rm nebula}).
\eqn\boveraneb$$
The striking agreement between equations \bovera\ and \boveraneb\ 
(which were derived independently) suggests that the ring may well be 
elliptical.

\chapter{Upper Limits}

Equation \alldata\ and its various modifications in \S\ 4 implicitly assume 
that the delay times are exactly equal to the excess distance divided by the 
speed of light.  That is, it is assumed that the fluorescent emission begins 
immediately when the EUV blast hits the ring gas.  It is possible that the 
emission is delayed while the gas recombines from highly ionized states or by 
some other unrecognized process.  However, such 
delays can only cause one to {\it overestimate} the physical size of the ring 
and thus, by equation \naivedist, overestimate the distance to the ring.  
Thus, all distance estimates should be regarded as upper limits rather than
measurements.  Since, the ring could plausibly be elliptical, we adopt the
overall upper limit from equation \excthp, and find $\mu_\lmc<18.44\pm 0.05$.

\chapter{Critique of Distance Estimate by Sonneborn et al.\ (1996)}

	As we discussed in \S\ 1, Sonneborn et al.\ (1996) used the same
data that we have analyzed here to derive a longer time to peak light
($t_\max = 399\,$days vs.\ $t_+ = 379\,$days) with larger errors (15 days vs.\
4.8 days), implying a longer distance to the supernova ring, also with larger
errors $(\mu=18.43\pm 0.10$ vs.\ $\mu=18.37\,\pm 0.04$).  We address four
questions related to this conflict:  Why did Sonneborn et al.\ (1996)
obtain a longer
time to peak light?  What significance does this longer time to maximum
have for the distance to the
ring?  Which curve gives a better fit to the data? 
Why are the Sonneborn et al.\ (1996) error estimates so much larger than ours?
The answers to these questions are directly related to the nature of the
method proposed by Gould (1995) for obtaining an {\it upper limit} to the 
ring distance.

	Why did Sonneborn et al.\ (1996) obtain a longer time to peak light?
Sonneborn et al.\ (1996) fit the data to a polynomial of order $n$, where
$n$ is a ``high'', but otherwise unspecified number.  In order to make a
concrete comparison of our work with theirs, we choose $n=9$ and fit to the
first 700 days of data.  This choice is made based on the local minimum in 
$\chi_{\rm eff}^2 = \chi^2 - (n+1)$ at $n=9$.  That is, there is no statistical
justification for incremental increases in the number of parameters beyond
$n=9$.
(For $n\geq 13$, $\chi_{\rm eff}^2$ begins to decline again, but the resulting
curves have the clear appearance of ``following the scatter''.)\ \
It is immediately clear from Figure \two\ that the prompt-response curve has
an earlier maximum than the polynomial because it has an 
{\it asymmetric cusp} which rises much
more steeply than it falls.  The polynomial, by contrast, is approximately
symmetric in the neighborhood of the peak.
	
	What significance does this longer time to maximum
have for the distance to the
ring?  In a word, ``none''.  It is certainly possible to draw curves that
come close to most data points and that have peaks at times $t_\max > t_+$.
However, the $t_\max$ from such curves are not related in any simple way
to the light-travel time to the far side of the ring and therefore cannot
be used to estimate the size of the ring (and hence its distance).  The
polynomial in Figure \two\ is an excellent example of such a curve.  It shows a
characteristic half-width at maximum
$$\Delta t = \biggl[{F(t_\max)\over F''(t_\max)}\biggr]^{1/2} = 125\,{\rm days},\eqn\deltat$$
where $F(t)$ is the polynomial.  Since the light curve from prompt-response
fluorescence is cuspy, the only {\it physical} way to produce such a broad 
peak is for
the response function of the gas to have a rise time $\sim 2\Delta t\sim
250\,$days.  If this were the case, the excess light travel time would
be a time $\sim t_\max-\Delta t\sim 275\,$days, not 400 days.

	Which curve gives a better fit to the data?  The prompt-response
curve is favored over the polynomial by more than $3\,\sigma$.  To determine
this, we normalize the error bars (as above) to make $\chi^2=59$ (the number
of degrees of freedom) for the best-fit prompt-response curve.  We then
find $\chi^2=63$ for the polynomial.  Since the polynomial fit has six more
free parameters, this implies $\Delta\chi^2=10$, 
corresponding to $3.1\,\sigma$.  However, we stress that this worse fit of
the polynomial curve has
no bearing on the problem of establishing an upper limit to the distance.
If the polynomial gave a better fit, it would not imply a larger
estimate for the ring size.  On the contrary, it would tend to indicate
a slow rise time for the fluorescence and hence an even smaller (and closer)
ring.

	Why are the Sonneborn et al.\ (1996) error estimates three times 
larger than ours (15 vs.\ 5 days)?  
First, we derived our estimate from N~III] and N~IV] together while they
used only N~III].  If we restrict our measurement to N~III], the error in
$t_+$ is 6.3 days.  Second, as we now show, Sonneborn et al.\ (1996) 
significantly overestimated their errors.  Consider a general linear function,
$F(t;a_1, ... ,a_n)= \sum_i a_i f_i(t)$, of which a polynomial is one
example.  Let $a_i^*$ be the best-fit parameters for the data.  The time of
maximum of this best-fit curve
is a solution of the equation $F'(t_\max;a_i^*) =0$, where the 
prime indicates differentiation with respect to time.  If the parameters
of the fit deviate by $\delta a_i$, then the solution of this equation will
change by $\delta t_\max \simeq -\sum_i\delta a_i f_i'(t_\max)/F''$. Thus,
the error in $t_\max$ is,
$$[{\rm var}(t_\max)]^{1/2} = \langle (\delta t_\max)^2\rangle^{1/2} =
{\Bigl[\sum_{i,j=1}^n c_{i j}f_i'(t_\max)f_j'(t_\max)\Bigr]^{1/2}\over
F''(t_\max)},\eqn\vartmax$$
where $c_{i j}\equiv \langle\delta a_i\delta a_j\rangle$ 
is the covariance matrix of the $a_i$ as derived by standard methods 
(e.g.\ Press et al.\ 1989) and as returned by linear fitting programs.  
That is, the error in $t_\max$
is the ratio of the error in the first derivative to the second 
derivative, both evaluated at $t_\max$.
Using equation \vartmax, we find
$[{\rm var}(t_\max)]^{1/2} = 6.3\,$days for the polynomial curve, i.e., 
the same as for the cuspy prompt-response curve.  

\chapter{Conclusions}

	The new reductions of the UV fluorescent emission-line data for
SN 1987A do not result in any major change in Gould's (1995) upper limit for
the distance modulus to the LMC under the assumption that the ring is
circular,
$\mu_\lmc<18.37\pm 0.04$, although the precision of the agreement between
this and equation \alldata\ is the result of an accidental cancellation of
two effects, each 0.02 mag.  However, new observational evidence for an
elliptical ring (previously considered to be an implausible hypothesis by
most workers in the field) raises the upper limit to $\mu_\lmc<18.44\pm 0.05$.
The result is still in strong conflict with
the recent Hipparcos-calibrated estimates of $\mu_\lmc\sim 18.65\pm 0.1$ based
on RR Lyraes and Cepheids (Reid 1997; Gratton et al.\ 1997; Feast \& 
Catchpole 1997).

	The robustness of the upper limit derives from the fact that the
delay times $t_\pm$ measure the light-travel time across the diameter of the
ring (eq.\ \naivedist).  
There are possible physical mechanisms that could retard these
times and so cause one to overestimate the distance.  However, there are none
(except superluminal motion) that could accelerate them.

	We have shown that the larger value of the time of maximum derived
by Sonneborn et al.\ (1996) is the result of choosing a non-physical
parameterization of the light curve.  Exactly the same criticism could be
made of the original determination by Panagia et al.\ (1991), and in fact
just such a criticism was made by Dwek \& Felten (1992) and elaborated upon
by Gould (1995).  We note that the simple 4-parameter physically-based light
curve of Gould (1995) fits the current data significantly better than a
10-parameter polynomial.

{\bf Acknowledgements}:  
We give special thanks to A.\ P.\ S.\ Crotts who analyzed
the geometry of the nebula in order to address specific questions raised by
this work and who provided these results in advance of publication.
We thank P.\ Popowski for stimulating discussions and D.\ Weinberg for
a careful reading of the manuscript.
This work was supported in part by grant AST 94-20746 from the 
NSF and in part by grant NAG5-3111 from NASA.

\Ref\be{Bessel, M.\ S., Freeman, K.\ C., \& Wood, P.\ R.\ 1986, ApJ, 310, 664}
\Ref\Ch{Crotts, A.\ P.\ S., \& Heathcote, S.\ R. 1991, Nature, 350, 683}
\Ref\Crotts{Crotts, A.\ P.\ S., Kunkel, W.\ E., \& Heathcote, S.\ R. 1995, 
ApJ, 438, 724}
\Ref\Dwek{Dwek, E.\ \& Felten, J.\ E.\ ApJ, 1992, 387, 551}
\Ref\fc{Feast M.\ W., \& Catchpole, R.\ W.\ 1977, MNRAS, in press}
\Ref\Gould{Gould, A.\ 1994, ApJ, 425, 51}
\Ref\Gouldb{Gould, A.\ 1995, ApJ, 452, 189}
\Ref\grat{Gratton, R.\ G, Pecci, F.\ F., Carretta, E., Clementini, G., 
Corsi, C., E., \& Lattanzi, M.\ 1997, ApJ, submitted (=astro-ph/9704150)}
\Ref\Jac{Jacoby, G.\ H., et al.\ 1992, PASP, 104, 599}
\Ref\Jak{Jakobsen, P., et al.\ 1991 ApJ 369, L63}
\Ref\Pana{Panagia, N., Gilmozzi, R., Macchetto, F., Adorf, H.-M. \&
Kirshner, R.\ P.\ 1991, ApJ, 380, L23}
\Ref\Pop{Popowski, P., \& Gould, A.\ 1997, ApJ, submitted}
\Ref\pr{Press, W.\ H., Flannery, B.\ P., Teukolsky, S.\ A.,
\& Vetterling, W.\ T.\ 1989, Numerical Recipes, 
(Cambridge: Cambridge Univ.\ Press)}
\Ref\plait{Plait, P., Lundqvist, P., Chevalier, R., \& Kirshner, R.\ 1995,
ApJ, 439, 730}
\Ref\redi{Reid, I.\ N.\ 1997, AJ, in press (=astro-ph/9704078)}
\Ref\xu{Xu, J., Crotts, A.\ P.\ S., \& Kunkel, W.\ E.\ 1995, ApJ, 451, 806}
\refout
\endpage
\end